\RequirePackage[l2tabu, orthodox]{nag}      
\documentclass[10pt,a4paper]{article}

\usepackage{amsmath}                        
\usepackage{amssymb}
\usepackage{graphicx}                       
\usepackage{microtype}                      
\usepackage{siunitx}                        
\usepackage{booktabs}                       

\usepackage{helvet}

\fontfamily{phv}\selectfont

\usepackage[english]{babel}
\usepackage[T1]{fontenc}                    
\usepackage{url}                            
\usepackage{makeidx}                        
\usepackage{multicol, caption}              
\usepackage{framed}                         
\usepackage{caption}                        
\usepackage{setspace}                       

\usepackage[usenames,dvipsnames]{color}
\usepackage{sectsty}
\usepackage{hyperref}                       

\usepackage[a4paper,left=3cm,top=2cm,bottom=3cm,right=3cm,ignoreheadfoot]
            {geometry}

\definecolor{gray90}{gray}{0.90}
\definecolor{gray80}{gray}{0.80}
\definecolor{gray70}{gray}{0.70}
\definecolor{gray60}{gray}{0.60}
\definecolor{gray50}{gray}{0.50}
\definecolor{gray40}{gray}{0.40}
\definecolor{gray30}{gray}{0.30}

\newcommand{\nfc}{8 }
\newcommand{\nfq}{61 }
\newcommand{\ngrs}{57 }
\newcommand{\njur}{4 }
\newcommand{\nindiv}{27 }
\newcommand{\ntotr}{$9.8 \times 10^9$}  

\newcommand{\fileSize}{60.8 megabytes } 

\newcommand{\kmer}{k-mer}
\newcommand{\fcda}{\textit{d24a}}
\newcommand{\fccoyr}{\textit{c0yr}}
\newcommand{\fcdor}{\textit{d10r}}
\newcommand{\fccog}{\textit{c0g9}}
\newcommand{\fccoyt}{\textit{c0yt}}
\newcommand{\fccuk}{\textit{c2uk}}
\newcommand{\fcdpd}{\textit{d1pd}}

\hypersetup{                            
    colorlinks=true,
    urlcolor=gray,
    linkcolor=gray50,
    citecolor=blue
}

\allsectionsfont{\sffamily\color{gray50}}

\makeatletter
\renewcommand{\maketitle}{\bgroup\setlength{\parindent}{0pt}
\begin{flushleft}
    \textbf{\@title}

    \vspace{3mm}
    \@author
\end{flushleft}\egroup
}
\makeatother

\title{{\Large Hierarchical clustering of DNA k-mer counts in RNA-seq
Fastq files reveals batch-effects}\\
\vspace{3mm}
\textnormal{
}}

\author{%
Wolfgang Kaisers$^{1}$, Holger Schwender$^{2}$ and Heiner Schaal$^{3}$\\
$^{1}$ \quad Heinrich Heine University, D\"{u}sseldorf\\
$^{2}$ \quad Mathematical Institute, Heinrich Heine University D\"{u}sseldorf\\
$^{3}$ \quad Institut fur Virologie, Heinrich Heine University D\"{u}sseldorf\\
\underline{$^{1}$kaisers@med.uni-duesseldorf.de}\\
}
\date{\today}

\makeindex

\begin{document}
\maketitle
\tableofcontents

\section{Abstract}

\textbf{Background:}
Batch effects, artificial sources of variation due to experimental design,
are a widespread phenomenon in high-throughput data.
Therefore, mechanisms for detection of batch effects are needed requiring
comparison of multiple samples.
We apply hierarchical clustering (HC) on DNA \kmer ~counts of multiple RNAseq
derived Fastq files.
Ideally, HC generated trees reflect experimental treatment groups and thus may
indicate experimental effects, but clustering of preparation groups indicates
the presence of batch effects.
In order to provide a simple applicable tool we implemented sequential analysis
of Fastq reads with low memory usage in an R package (seqTools) available on
Bioconductor.
DNA \kmer ~counts were analysed on \nfq Fastq files containing RNAseq data
from two cell types (dermal fibroblasts and Jurkat cells) sequenced on \nfc
different Illumina Flowcells.
%
\textbf{Results:}
%
Pairwise comparison of all Flowcells with hierarchical clustering revealed
strong Flowcell based tree separation in 6 (21 \%) and detectable Flowcell
based clustering in 17 (60.7 \%) of 28 Flowcell comparisons.
In our samples, batch effects were also present in reads mapped to the human
genome.
Filtering reads for high quality (Phred $>$30) did not remove the batch effects.
%

\textbf{Conclusions}
%
Hierarchical clustering of DNA \kmer ~counts provides a quality
criterion and an unspecific diagnostic tool for RNAseq experiments.

\section{Introduction}

In hierarchical clustering (HC), different entities are located in bi-parting
trees according to their pairwise similarity (quantified by a distance measure).
Although the trees provide no absolute measure, accumulation of biological or
technical related samples in different sub-trees indicates sample similarities
which possibly influence downstream analysis.
The interpretation of dendrograms requires (at least to some degree) subjective
validation.\\
DNA \kmer 's are short sequence patterns in DNA sequence of length k.
Counting of DNA \kmer's in Fastq files results in count vectors of length $4^k$
to which distance measures (for example the Canberra distance) can be applied.
\\
Transcriptome sequencing data is prone to disturbing effects resulting from
experimental design for example ozone levels \cite{pmid14632079}, random hexamer
priming \cite{RandHexamer}, GC-content \cite{pmid22177264}, and transcript
length \cite{pmid21252076}.
Batch effects, variation due to experimental design, is a prevalent phenomenon
and clustering according to surrogate values (processing date or batch) is a
common approach for recognition of underlying sources of variation
\cite{BatchIri}.\\
We apply hierarchical clustering to DNA \kmer ~content in Fastq files (HcKmer)
and analyse transcriptome sequencing data of 61 samples.
Sample clustering from raw Fastq files, (Phred) quality filtered sequences and
from aligned sequences are analysed.
A putative effect on analysis results of differential gene expression is
explored.
Additionally, the sensitivity of HcKmer is examined in a simulation study.
%
%
\subsection{Algorithmic framework}

For a DNA sequence of length k, $4^k$ different sequence motifs (\kmer s) exist
each defining a category for which occurrences can be counted.
The value of k is usually chosen in the range of 5 to 9 (resulting in 1,024 to
262,144 different \kmer s) as shorter motifs impair separation capabilities and
longer motives increase computational demands.\\
The algorithms for performing HcKmer are available in R package
\textit{seqTools} on Bioconductor \cite{seqTools}.
DNA \kmer ~counts are collected with other quality measures (for example Phred
scores and GC content) from Fastq files.
Data from multiple samples (for example from whole Illumina Flowcells) can be
collected at once into merge-able objects.
Combined objects then serve as input for hierarchical clustering.
The Canberra distance, defined as
\begin{align}\label{candist}
  f_{cb}(x_i,y_i) &=& \frac{|x_i-y_i|}{|x_i|+|y_i|}\\
  d(x,y)          &=& \sum_{i=1}^k f_{cb}(x_i,y_i),
  \quad (x_i)_{i=1,\ldots,k}, (y_i)_{i=1,\ldots,k} \in \mathbb{R}^k,
\end{align}
is used as distance measure between samples.
Total read numbers ($\sum_{i=1}^m x_i$) are scaled to a common value
in order to compensate a systematic offset caused by different sequencing
depth's (thereby increasing $|x_i-y_i|$ values).
Subsequent cluster-analysis was calculated using the \textit{hclust} function
(CRAN \textit{stats} package) \cite{CRAN}.

\subsection{Sample preparation, sequencing and alignment}

In this study, transcriptome data from \nfq samples is analysed.
Therefrom, \ngrs samples of human short term cultured human dermal fibroblasts
were obtained from \nindiv human healthy individuals and sequenced for a study
on human ageing \cite{pmid28475575}.
Collection and processing of dermal samples from donors was approved by the
Ethical Committee of the Medical Faculty of the University of D\"{u}sseldorf
(\# 3361) in 2011.
The Fastq files from these samples are available under ArrayExpress accession
E-MTAB-4652 (ENA study ERP015294).
Also, \njur samples from cultured human Jurkat cell lines were sequenced
for a study on HIV infection.\\
Sample preparation and sequencing has been described elsewhere
\cite{pmid28475575}.
In short, cellular mRNA was amplificated on \nfc Illumina Flowcells (v1.5)
and sequenced on a Illumina HiSeq 2000 sequencer.
From each lane, the resulting 101-nucleotide sequence reads were converted
to Fastq by CASAVA 1.8.2.
A Fastq file contained in average $162.2 \times 10^6$ reads.
In total the \nfq Fastq files contained \ntotr ~reads.\\
Subsequent alignments were calculated on unprocessed Fastq files with
\textit{TopHat} (v 2.0.14) using human GRCh38 assembly.\\
In order to compare batch effects between raw Fastq files and mapped reads,
BAM file content was transformed back into Fastq (using \textit{bam2fastq}
from CRAN package rbamtools \cite{pmid25563331}).
All DNA \kmer ~counts were collected using k=9 ($4^9 =$ 262,144 DNA motifs),
except for the simulation study (k=6 is used for HcKmer).\\
Differential gene expression analysis was performed using Quasi-likelihood
F-Tests from the edgeR (3.12.0) framework \cite{pmid19910308}.
Genes with a reported FDR $<0.1$ were considered to be significantly
differential expressed.
The same analytic procedure performed on the whole dataset resulted in
\textit{no} significantly differential expressed gene \cite{pmid28475575}.
%
%

\section{Results}

\subsection{Data collection}

The \nfq samples had been sequenced on \nfc Flowcells.
Data from each Flowcell was collected into one data-set.
Processing of \ntotr ~reads took 8.96 hours ($3.04 \times 10^5 $ reads/second)
in a single thread with approximately 1 Gigabyte working memory consumption.
The saved raw data was \fileSize in size.
\begin{framed}
    \begin{center}
        \includegraphics[scale=0.5]{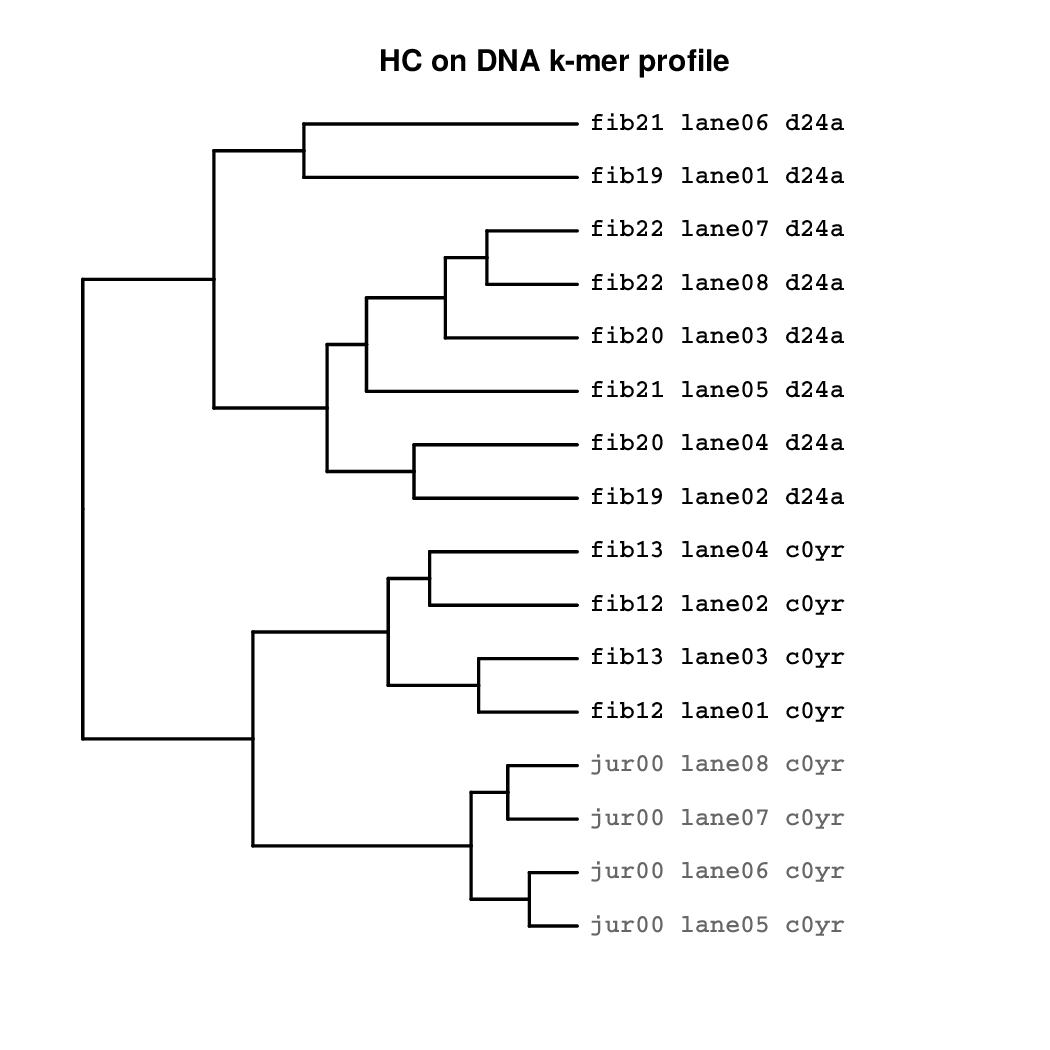}
    \end{center}
    \captionof{figure}{\small
        \textbf{Strong batch effect in comparison of a Flowcell pair.}
        Leaf labels denote cell type (fib, jur) and number of individual,
        lane number and Flowcell label (\fcda ~and \fccoyr).
        Samples from Jurkat cells are highlighted in grey.
        The tree clearly separates Flowcell \fcda ~and \fccoyr ~although Flowcell
        \fccoyr ~contains two different cell types.}
    \label{fig:strong_batch}
\end{framed}
%
%
\subsection{Identification of batch effects.}
%
\figurename ~\ref{fig:strong_batch} shows a dendrogram where 16 samples from two
Flowcells (\fcda ~and \fccoyr) and two cell types (dermal fibroblasts and Jurkat
cells) are analysed using HcKmer.
The partition tree consists of two equally sized sub-trees.
Each of them only contains samples sequenced on one Flowcell although on
Flowcell \fccoyr, two different cell types are present.
Thus, DNA motif dissimilarity is greater between Flowcells \fcda ~and \fccoyr
~than between fibroblasts and Jurkat cells.
\figurename ~\ref{fig:phred_median_d24a_c0yr}, indicates that median Phred
scores in sequences from both Flowcells are sufficiently high.
%
\begin{framed}
    \begin{center}
        \includegraphics[scale=0.5]{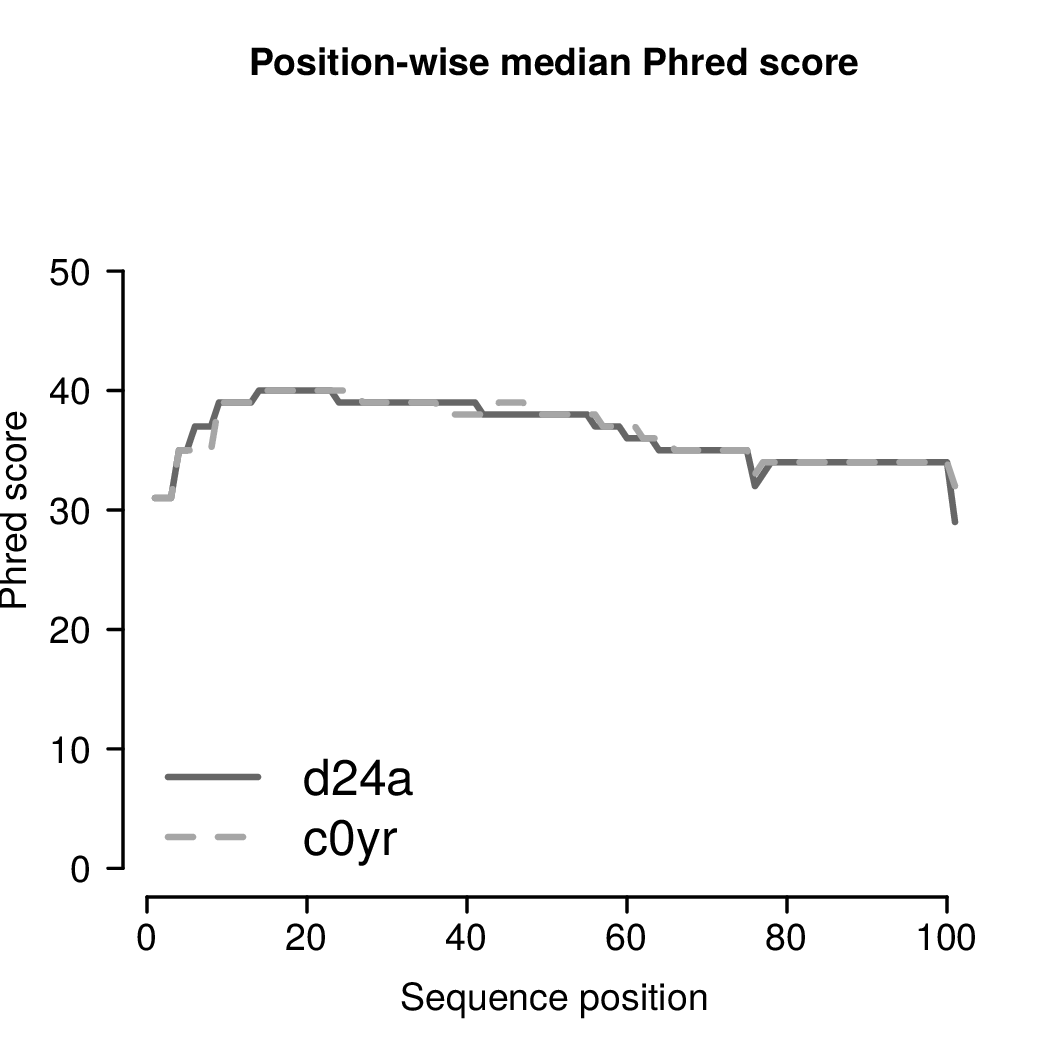}
    \end{center}
    \captionof{figure}{\small
        \textbf{Median Phred score values.}
        For each read position, median Phred scores are shown for samples
        sequenced on Flowcells \fcda ~and \fccoyr.
        All median Phred scores are $>$ 28.}
    \label{fig:phred_median_d24a_c0yr}
\end{framed}
%
\subsection{Filtering for sufficient Phred scores}

In order to explore, whether the observed dissimilarity can be removed, reads
containing at least one Phred score $<$ 30 were discarded.
The filter was applied to Fastq files sequenced on \fcda ~and \fccoyr ~Flowcells.
From Flowcell \fcda, 14.3 \% of reads in Fastq files and from Flowcell \fccoyr,
15.1 \% of reads in Fastq files were excluded thereby.
Finally, the filtered reads were re-analysed.\\
In the HC-tree, the position of one Flowcell (\texttt{fib13 lane03 c0yr})
changed to the opposite sub-tree but still, three fibroblast samples cluster
together with the Jurkat samples (\figurename ~\ref{fig:hc_filtered_d24a_c0yr}).
\begin{framed}
    \begin{center}
        \includegraphics[scale=0.5]{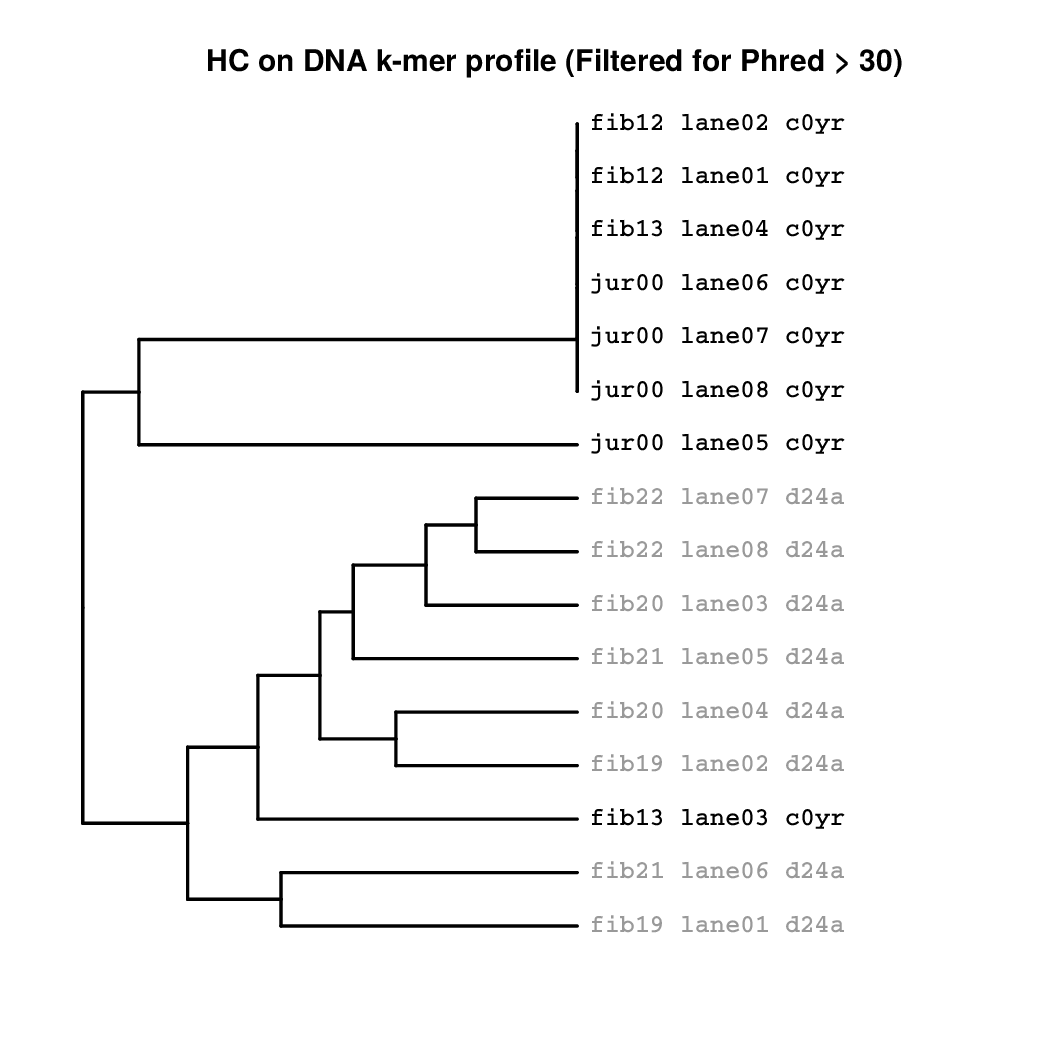}
    \end{center}
    \captionof{figure}{\small
            \textbf{Clustering of Fastq files containing filtered reads.}
            All reads containing Phred scores $<$ 30 had been discarded before
            HcKmer analysis.
            On top level clade, the Jurkat cell samples still exclusively
            cluster together with samples from the same Flowcell.}
    \label{fig:hc_filtered_d24a_c0yr}
\end{framed}
Thus, filtering based on Phred scores induced only a minor change in cluster
formation.

\subsection{Detection of experimental effects}

Likewise batch effects, clustering of experimental groups may be indicative for
experimental effects.
The analysed fibroblast samples had been collected from donors in different
age groups (Young: 19 to 25 years, Middle: 36 to 45 years, Old: 60
to 66 years) \cite{pmid28475575}.
\figurename ~\ref{fig:indiv_clust} shows a dendrogram, where all samples
from donors aged $< 60$ years are located within the same subtree while
two samples sequenced on the same flowcell (\fccoyt) are located outside
this subtree.
As appearance of these clusters may indicate experimental effects,
putative experimental sources should be explored.
As DNA \kmer ~content is an unspecific indication, potential causes are not
restricted to differential gene expression but may include differential
(alternative) splicing events.
\begin{framed}
    \begin{center}
        \includegraphics[scale=0.5]{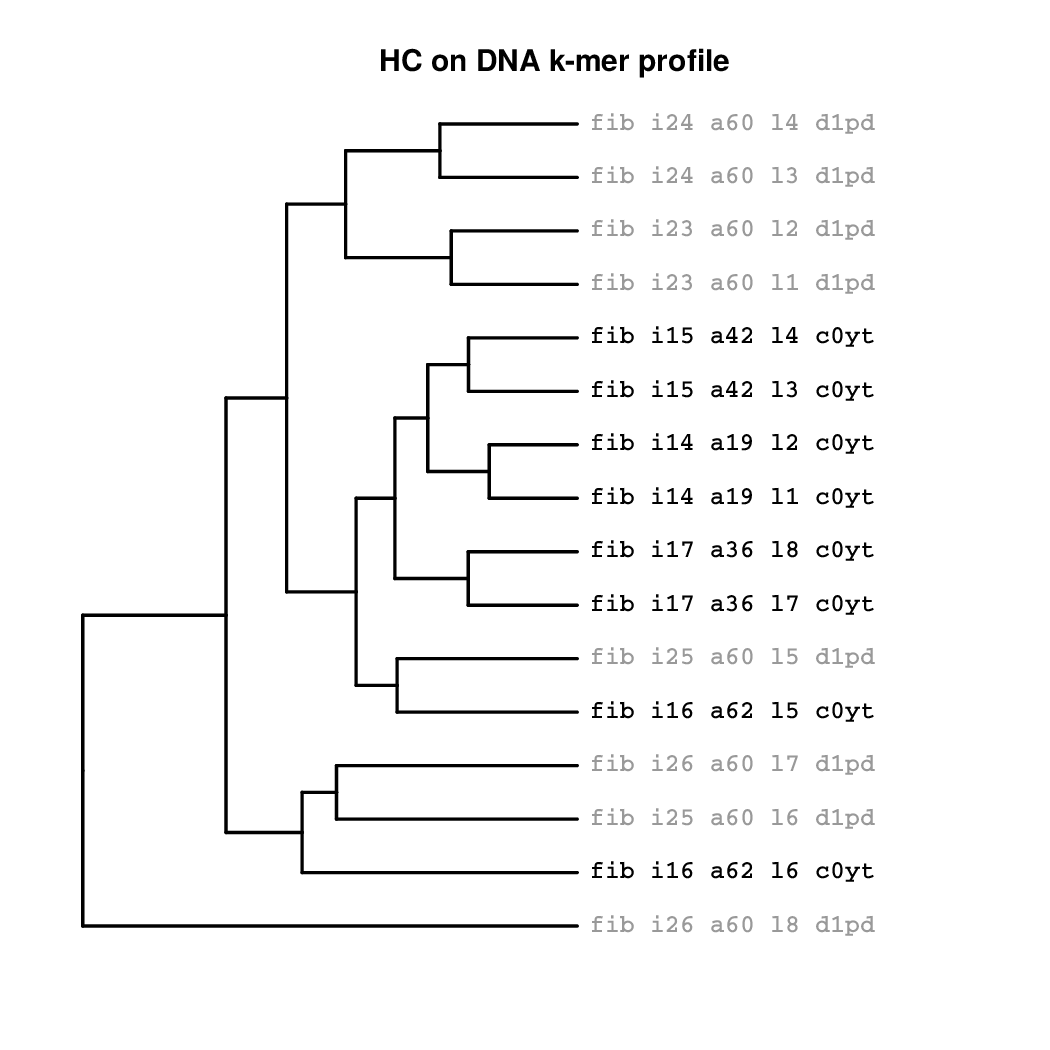}
    \end{center}
    \captionof{figure}{\small
        \textbf{Clustering according to experimental groups.}
        Leaf labels denote cell type (fib), id of sample donor,
        age sample donor, Flowcell lane and Flowcell label
        (\texttt{fib i24 a60 l4 d1pd} refers
        to fibroblast, donor-id 24, donor age 60 years and Flowcell \fcdpd).
        Samples from Flowcell \fcdpd ~are highlighted in gray.
        All samples where donor age is $<45$ years cluster in a separate
        sub-tree.
        }
    \label{fig:indiv_clust}
\end{framed}
%

\subsection{HcKmer on sequences aligned to the human genome}

Fastq reads causative for HcKmer tree separation may not match to the
reference genome and thus would be filtered out by alignment.
In order to evaluate this eventuality, clustering of raw Fastq files and
mapped reads are compared.
Pairs of Flowcells containing 8 fibroblast samples (\fcda, \fcdor, \fccog,
\fccoyt, \fccuk, \fcdpd) are analysed by HcKmer on raw reads and on mapped
reads.
An example is shown in \figurename ~\ref{fig:batch_mapped} where raw and
mapped reads cluster in similar patterns.
In both trees, the size of the largest subtree containing samples from only one
Flowcell is 7.
The mean sizes of these subtrees in all 15 possible pairs is 7.21 in raw reads
and 6.57 in mapped reads.
Mapping thus reduces Flowcell cluster sizes by 8.9 \% but still, clusters of
size $\geq 6$ are prevalent.
Detailed results are shown in supplemental data.

\begin{framed}
    \begin{center}
        \includegraphics[scale=0.5]{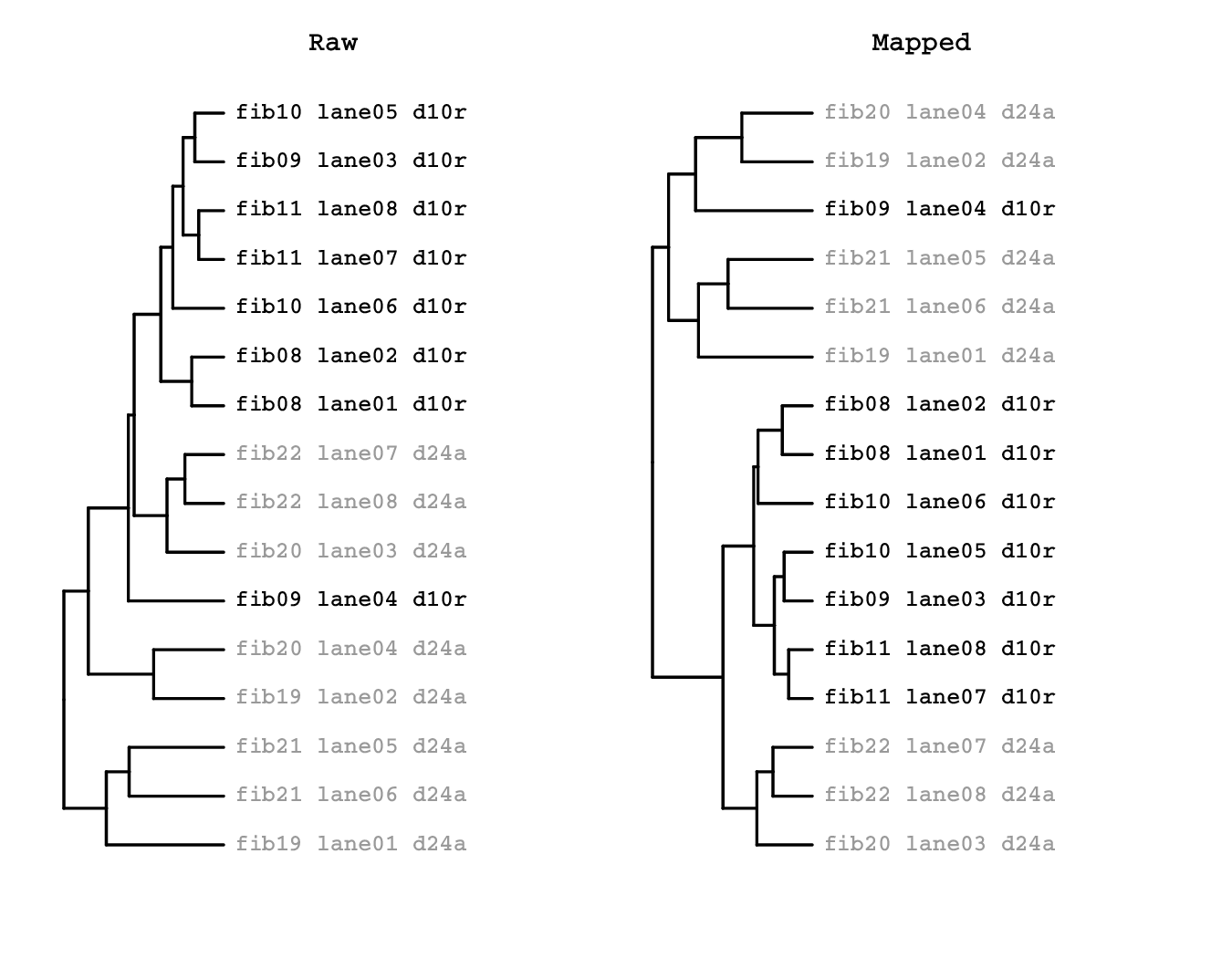}
    \end{center}
    \captionof{figure}{\small
        \textbf{HcKmer on raw and mapped reads}
        Fastq files containing mapped reads were constructed from raw reads
        by alignment using TopHat followed by extraction of reads from
        BAM files using \textit{reader2fastq} (rbamtools) function.
        HcKmer on raw and mapped reads was calculated on Fastq files from the
        same Flowcell pairs.
        Raw reads consist of unmapped and mapped reads.
        \textbf{Left :} HcKmer on raw reads. \textbf{Right: } Mapped reads.
        Leaf labels denote cell type (fib), lane number, and Flowcell label
        (\fcda ~and \fcdor).
        All but one samples from Flowcell \fcdor ~(dark grey) cluster within
        a separate sub-tree.
        Raw and mapped reads show similar clustering characteristics.
        }
    \label{fig:batch_mapped}
\end{framed}

\subsection{Separation sensitivity on simulated data}

The degree of sequence dissimilarity required to produce HcKmer trees with
top level separation of two sample groups (e.g. Flowcells) is determined in
simulation where Fastq files with random DNA are analysed.
A pure random group is compared to a group in which a variable percentage of
random DNA reads is contaminated with (one ore multiple) fixed DNA 6-mers.\\
Clustering of groups is quantified using a score (Contralaterality Score, CS).
A decrease of CS from 41.1\% (mean value for pure random sequences) to 0\% is
regarded as indicative for clustering of groups and in the setting of the
simulation, a CS $<$ 10 \% can considered to be statistical significant
(p $<$ 0.05).\\
\figurename ~\ref{fig:sim_single_cont} shows separation capabilities of HcKmer
for contamination of 0 - 6 \% of Fastq reads with one fixed DNA 6-mer.
The results indicate that with a contamination of 4 \% of Fastq reads,
significant sample separation reaches a power of 80 \%.
Details of simulation, definition of CS and results are shown in supplemental
data.\\
\begin{framed}
    \begin{center}
        \includegraphics[scale=0.5]{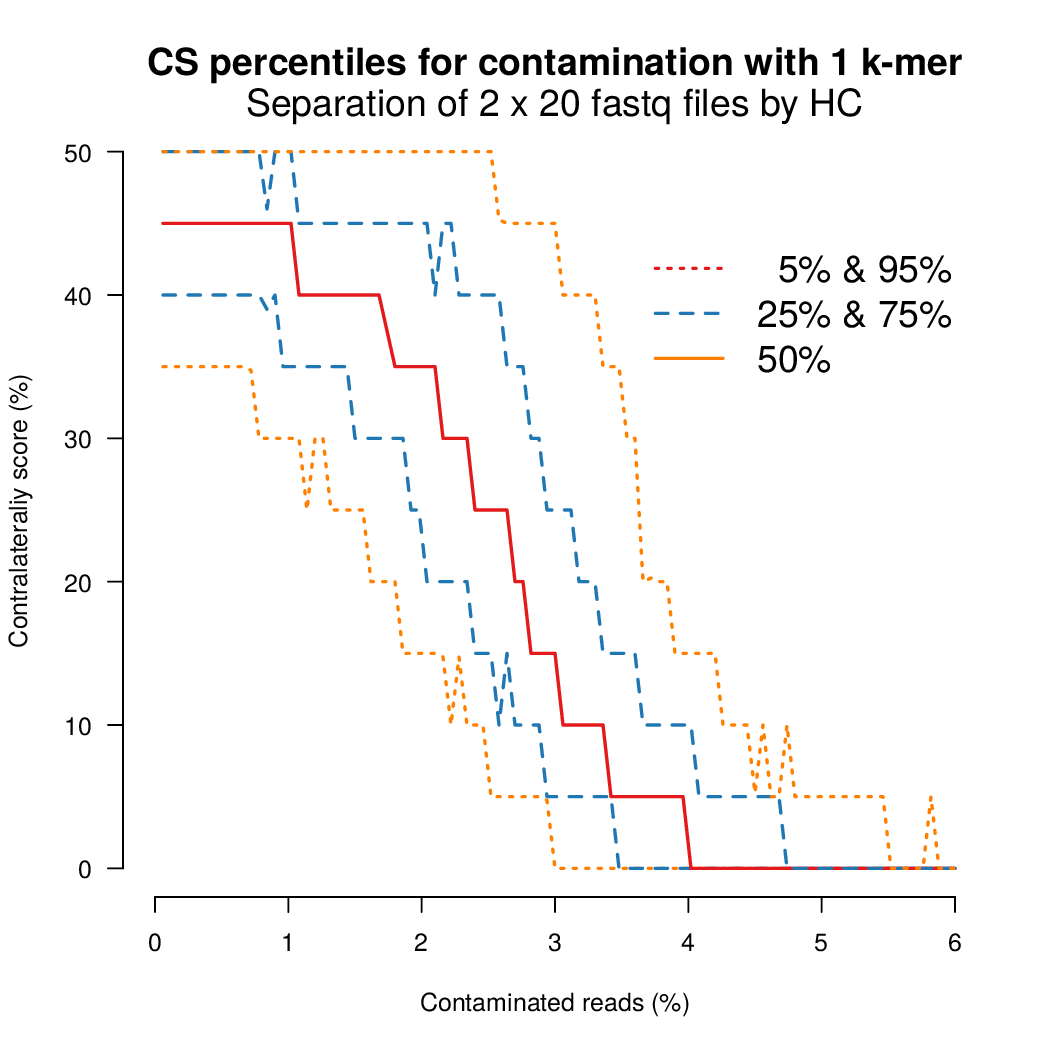}
    \end{center}
    \captionof{figure}{\small
        \textbf{Separation sensitivity on simulated data.}
        HcKmer was performed on Fastq files containing simulated DNA sequence.
        Percentiles for Contralaterality score (CS) are given for variable
        percentages of contamination with a single fixed DNA 6-mer.
        CS quantifies the presence of the minor present group in the first
        half of the HC-derived group labels.
        CS values \textless 10 \% are considered to be statistically
        significant (p \textless 0.05).}
    \label{fig:sim_single_cont}
\end{framed}

\subsection{Prevalence of batch effects in RNAseq data}

The prevalence of detectable batch effects is analysed on the whole set of 61
samples sequenced on 8 Flowcells.
Pairwise comparison of 8 Flowcells results in 28 pairs.
All Flowcell pairs are analysed for presence of batch effects using
HcKmer and a semi-quantitative score ranging from strong batch effect (b1a =
top-level separation of Flowcells; shown in \figurename ~\ref{fig:strong_batch})
and detectable batch effect (b1a, b1b, b2a or b2b) to absence of batch effects
(es).
From the analysed Flowcell pairs, 6 (21.4 \%) show strong batch effects and
17 (60.7 \%) show detectable batch effects.
Thus, batch effects are present in a considerable fraction of Flowcell pairs.
The definitions and details on analysis are shown in supplemental material.\\

\subsection{Influence of batch effects on false discovery rate}

Clustering according to sample preparation batches potentially influences
differentially Expressed Gene (DEG) analysis in RNAseq data.
To address this question, results from DEG analysis are compared between
Flowcells with different HcKmer dissimilarity using a two-way ANOVA.
Classification as batch effect (b1 or b2 vs. es) is significantly associated
with increased number of differentially expressed (DE) genes
(see supplemental material for details).
The ANOVA predicts 3.848 DE genes for b1a dissimilarity and 695 DE genes
for idf/es (no batch effect detectable).
Thus, increased number of false positives in DEG analysis are found when
batch effects are identified by HcKmer.

\subsection{K-mer spectrum responsible for tree separation}

Standard quality control tools (for example FastQC) only report a small number
of over-represented \kmer s which may not represent a sufficient sample for
explanation of batch effects diagnosed by HcKmer.
We therefore estimate how many \kmer s are needed in order to evoke the
observed batch effects using an example.\\
From the simulation data on separation sensitivity, it is assumed that
$\approx 3 \%$ contamination (where the median CS falls below 12.5 \% in
\figurename ~\ref{fig:sim_single_cont}) is required in order to produce a
strong batch effect (b1a).
Two pairs of Flowcells are selected, one with complete tree separation due
to batch effect (Flowcells \fcda ~and \fccoyr) and one pair of Flowcells without
tree separation due to batch effect (Flowcells \fcda ~and \fcdpd).
For each \kmer, the logarithmised and normalised sum of \kmer ~counts on whole
Flowcells are calculated and compared for both pairs and are shown in
\figurename ~\ref{fig:SmoothScatLogCkmer}).
In the example shown in left panel (\fcda /\fccoyr ~comparison), differences
from the 3,678 \kmer s with the largest difference in \kmer ~count need to be
accumulated in order to attain a 3 \% contamination rate.\\
Due to this large number it is unlikely, that strong batch effects (b1a)
diagnosed by HcKmer can be identified by inspecting few \kmer s and that they
can be eliminated by removing reads containing a small set of selected
\kmer s.
%
%
\begin{center}
\fbox{
    \begin{minipage}{.97\textwidth}
        \begin{center}
            \includegraphics[scale=0.5]{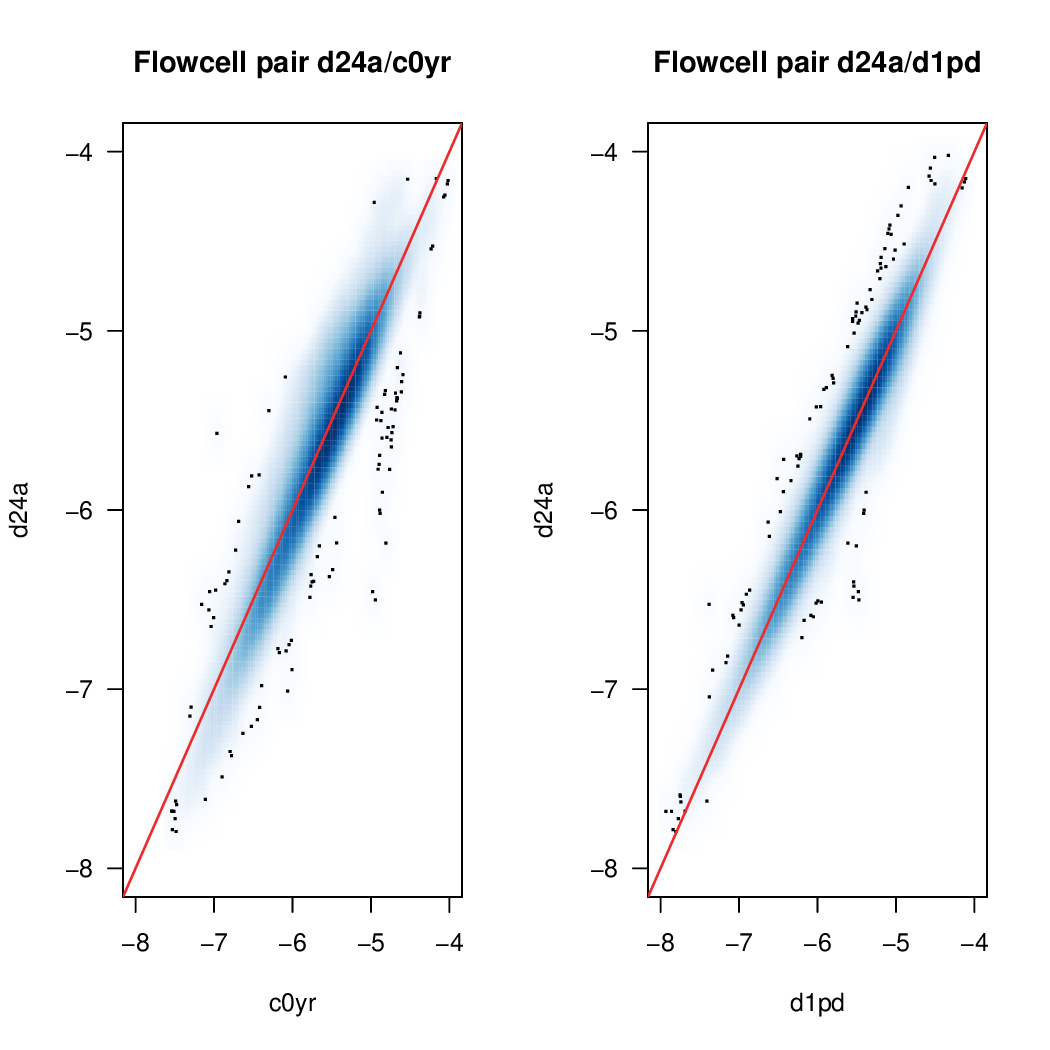}
        \end{center}
        \captionof{figure}{K-mer spectrum responsible for tree separation.}
        \label{fig:SmoothScatLogCkmer}
        {\footnotesize
        \begin{onehalfspace}
            Comparison of K-mer counts in two Flowcell pairs using
            density scatter plots:
            All axes represent normalised and $\log_{10}$ transformed \kmer
            ~counts on a whole Flowcell (each 8 Fastq files).
            The diagonal (red line) indicates equal normalised counted numbers
            for \kmer s in both Flowcells.
            \textbf{Left panel}: Comparison of \kmer ~counts on Flowcells
            \fcda /\fccoyr (strong batch effect (=b1a) identified by HcKmer).
            \textbf{Right panel}: Comparison of \kmer ~counts on Flowcells
            \fcda /\fcdpd ~(no batch effect (=es) identified by HcKmer).
            \textbf{Result:} The k-mer count differences are larger for
            \fcda /\fccoyr ~(mean= 0.30, sd= 0.12) than for \fcda/\fcdpd
            ~(mean=0.01, sd=0.07).
            The HcKmer diagnosed difference of sample similarity between the
            Flowcell pairs is due to larger deviation of \kmer ~counts from the
            diagonal for a broad variety of \kmer s (thus not generated by a
            small group \kmer s with large
            deviation).
        \end{onehalfspace}
        }
    \end{minipage}
}
\end{center}

%

\section{Discussion}

Differences identified by HcKmer should initiate exploration of underlying
effects because comparison of raw and mapped reads as well as differential
expression analysis on our samples indicate, potentially influential effects on
downstream analysis.\\
Clustering of samples according to biological or experimental entities is an
unspecific criterion as different K-mer spectra may be caused by differential
gene expression, differing splicing patterns as well as the mentioned disturbing
factors (for example random hexamer priming, GC-content or transcript length).
Thus etiologic factors may not be apparent.\\
Analysis of our samples shows, that clustering of preparation batches
may be prevalent (21 \% - 61 \%) and that resulting disturbances may not be
removed by Phred based filters or by alignment to a genome.
Also, batch effects potentially affect results of DEG analysis.\\

Due to analysis results of differential gene expression on our fibroblast
samples \cite{pmid28475575}, we assume that there are no consistent differences
between different groups and thus, GC-content and transcript length are
unlikely to evoke the observed batch effects.
Also, observed sample dissimilarities seem to result from a larger number of
small differences in \kmer ~counts which can not be diagnosed by identification
of few over- or underrepresented \kmer s.
Based on the shown results, our group decided to discard data from samples
sequenced on two other Flowcells (data not shown here) exhibiting strong
batch effects when compared with all Flowcells in the shown analysis.\\
HcKmer provides no quantitative measure for similarity so that subjective
judgement is required.
Also, a minimum of four samples per group are required for assessment.
But, as HcKmer can easily be applied (using \textit{seqTools}) and HcKmer
can identify influential effects not detected by other QC procedures, it appears
advisable to include HcKmer into analysis standards.

\section{Conclusions}

HcKmer provides an unprejudiced view onto the raw data produced by a sequencing
experiment with the capability of detecting unwanted variation which is
prevalent and potentially influential.
Experimental designs allowing HcKmer analysis (for example by sequencing samples
together on a defined set of Flowcells and to avoid multiplexing) thus are
favourable.
Based on contrasts identified by HcKmer, further exploration of results from
sequencing experiments as well as exclusion of contaminating samples may be
reasonable.

\subsubsection{List of abbreviations}

\begin{tabular}{@{}ll}
HC      & Hierarchical clustering\\
HcKmer  & Hierarchical clustering of DNA k-mer counts\\
DEG     & Differential expressed gene(s)\\
QC      & Quality control\\
\end{tabular}

\subsubsection{Ethics approval}
The study was approved by the Ethical Committee of the Medical Faculty of the
University of D\"{u}sseldorf (\# 3360) in 2011.

\subsubsection{Availability of data and materials}
The used software (seqTools) is available from the Bioconductor web site.
The raw Fastq files for fibroblast samples are available from ArrayExpress
under accession E-MTAB-4652 (ENA study ERP015294).
%

\subsubsection{Competing interests}
The authors declare that they have no competing interests.





\printindex
\end{document}